\pdfoutput=1
\documentclass[twocolumn,amsmath,longbibliography,amssymb,superscriptaddress]{revtex4-1}
\usepackage[pdftex]{graphics}
\usepackage{graphicx}
\graphicspath{{figures/}}
\usepackage{hyperref}
\usepackage{xcolor}
\usepackage{physics}
\usepackage{subfig}
\usepackage{bm}
\usepackage{caption}
\usepackage{amsmath} 
\usepackage{mathtools}

\newcommand{\tpo}{\tilde{\mathcal{P}}_{\rm o}}
\newcommand{\brapsio}[1]{\bra{\tilde{\Psi}_{{\rm o}}^#1}}
\newcommand{\ketpsio}[1]{\ket{\tilde{\Psi}_{{\rm o}}^#1}}
	
\begin{document}
		
\title{On the relation of the entanglement spectrum to the bulk polarization}
\author{Carlos Ortega-Taberner}
\affiliation{Department of Physics, Stockholm University, AlbaNova University Center, SE-106 91 Stockholm, Sweden}
\affiliation{Nordita, KTH Royal Institute of Technology and Stockholm University, SE-106 91 Stockholm, Sweden}

\author{Maria Hermanns}
\affiliation{Department of Physics, Stockholm University, AlbaNova University Center, SE-106 91 Stockholm, Sweden}
\affiliation{Nordita, KTH Royal Institute of Technology and Stockholm University, SE-106 91 Stockholm, Sweden}
\date{\today}
\begin{abstract}
The bulk polarization is a $\mathbb{Z}_2$ topological invariant characterizing non-interacting systems in one dimension with chiral or particle-hole symmetries. We show that the bulk polarization can always be determined from the single-particle entanglement spectrum, even in the absence of symmetries that quantize it. In the symmetric case, the known relation between the bulk polarization and the number of virtual topological edge modes is recovered. We use the bulk polarization to compute Chern numbers in 1D and 2D, which illuminates their known relation to the entanglement spectrum. Furthermore we discuss an alternative bulk polarization that can carry more information about the surface spectrum than the conventional one and can simplify the calculation of Chern numbers. 
\end{abstract}

\maketitle

\section{Introduction}
Topological phases of matter have attracted a lot of attention during the last decades, not the least because a large variety of relevant systems have been realized experimentally. 
The early focus was mainly on \emph{topologically ordered} systems \cite{wenbook}, where interaction effects are crucial for stabilizing the phases. 
The most notable examples are the fractional quantum Hall liquids~\cite{Tsui1982} and quantum spin liquids~\cite{Balents2010spin}. 
However, since 2005~\cite{kane2005quantum, roy2009topological} the focus has shifted to  \emph{symmetry-protected topological phases} (SPT) where symmetries are necessary to protect the topological phases and determine which distinct topological phases can be realized for a given dimensionality. These can be implemented as free-theories and can be characterized in terms of topological invariants~\cite{ryu2010topological}. 

Entanglement has played an important role in the understanding and characterization of topological systems \cite{Ludwig2015,Wen2017}. One can distinguish two types of states depending on their entanglement. Short range entangled states can be continuously transformed into a direct product state, while long range entangled states cannot. The latter correspond to topologically ordered states. Certain short range entangled states cannot be continuously transformed between themselves unless certain symmetries are broken. These are the SPT phases that we focus on in this paper. 

There are different tools based on entanglement that have been used to characterize topological phases. A very efficient one is the entanglement entropy~\cite{srednicki1993entropy}, which allows one to determine the total quantum dimension of the underlying topological quantum field theory~\cite{Kitaev2006topological, Levin2006detecting}. However, it can only be used for topologically ordered phases and it cannot uniquely characterize the topological phase at hand.
Another, closely related, tool is the entanglement spectrum (ES), originally introduced for fractional quantum Hall systems~\cite{Li2008entanglement}. It provides information about the edge spectrum and has proven useful for other topologically ordered phases such as fractional Chern insulators~\cite{Regnault2011fractional} and certain quantum spin liquids~\cite{yao2010entanglement}.

For \emph{non-interacting} topological insulators  and superconductors the ES for (gapped) periodic systems can be computed very efficiently, using methods developed by Peschel and others~\cite{Peschel2003}. The ES in these systems is equivalent to the flat-band energy spectrum of the corresponding system with open boundaries~\cite{Fidkowski2010entanglement}. The same correspondence was also found for closely related gapless systems~\cite{Matern2018entanglement}. However, even for non-interacting systems, it is unclear which information (beyond the protected `edge' spectrum) is encoded in the ES. The aim of this paper is to show that other physical properties are also encoded in the ES, in particular the bulk polarization.

The bulk, or macroscopic, polarization is a fundamental concept in physics, primarily used to describe the response of matter to electric fields. The modern theory of polarization \cite{Resta1992,KingSmith1993,Vanderbilt1993,Resta1997} related the bulk polarization to a geometric phase, which is nothing but the Zak phase for translationally invariant systems.  Due to this, it has also found its way into topological physics \cite{Kudin2007}. In certain symmetry classes the bulk polarization is quantized and serves as a $\mathbb{Z}_2$ topological invariant, where it is known to be related to certain feature of the ES \cite{Ryu2006}. Because of its relation to a geometric phase, the bulk polarization is also related to other topological invariants in higher dimensions such as the Chern number.

We show that there is much more information encoded in the \emph{single-particle} ES than was previously known. In particular, we show how the bulk polarization can be decoded from the single-particle ES, even when it is not quantized by the symmetries. This is a general property of non-interacting 1D gapped systems. We also apply this method to compute Chern numbers directly from the single-particle ES. The differences in the bulk polarization, which can be any real number, are the relevant quantities to study. They can, however, be difficult to compute from the bulk polarization itself, as it is only defined modulo $1$. We show how, using our method, one can define an alternative bulk polarization, computed by using open boundary conditions, which is continuous in $\mathbb{R}$ for gapped paths in parameter space, simplifying the calculation of polarization differences. The bulk polarization defined this way can give more information about the edge spectrum than the conventional one, and can be used to simplify the computation of Chern numbers. We also discuss the relation between this alternative bulk polarization and another topological invariant known as the trace index \cite{Alexandrinata2011}. 

The relation between the ES and the bulk polarization has already been studied. 
In Ref.~\cite{Zaletel2014}, the authors show that the Zak phase can be computed from the Schmidt decomposition of a translationally invariant, infinite chain. The latter is related to the ES.
This suggests that, for non-interacting systems, there should be a direct relation between the Zak phase and the single-particle ES, which is substantially easier to compute than the ES. A particular limit of our result was derived for a fully dimerized SSH chain~\cite{Ryu2006}. 
The similarity between the behavior of the Zak phase and the ES of certain Chern insulators, which was observed in Ref.~\cite{Huang2012,Huang2012-2}, is also explained by our results.

\emph{Outline of the paper}: In section II we introduce the 1D model used throughout the paper and we examine its phase diagram. In section III we discuss the bulk polarization; the different ways one can define it and its relation to the geometric phases. The ES is introduced in section IV, where we also set our notation. In section V we present our method and show how we can reproduce the bulk polarization for systems with and without translational invariance. In section VI we introduce an alternative bulk polarization constructed from the ES and use it to compute Chern numbers in 1D and 2D. Finally in section VII we conclude and discuss possible extensions of our work.

\section{The model}
To illustrate our results we consider a model in the BDI class~\cite{Song2014}. Although simple, it supports a rich phase diagram containing topological phases with winding numbers $\nu =0,1$ and $2$. 
In order to show that our results are not limited to this particular symmetry class~\cite{ryu2010topological}, we also include two symmetry breaking terms. 
The Hamiltonian is then
\begin{align}
H =& \sum_{i\alpha,j\beta} c_{i\alpha}^\dagger H_{ij,\alpha \beta} c_{j\beta},
\end{align}
where 
\begin{align}
H_{ij} =& (m \sigma_x + \kappa \sigma_z)\delta_{ij}  + \frac{1}{2i}\kappa'\sigma_z (\delta_{i-j,1}-\delta_{i-j,-1})\nonumber\\
&+ \frac{1}{2} t \left[(\sigma_x + i \sigma_y)\delta_{i,j+1} + (\sigma_x - i \sigma_y) \delta_{i,j-1} \right] \nonumber\\
&+  \frac{1}{2} t' \left[(\sigma_x + i \sigma_y)\delta_{i,j+2} + (\sigma_x - i \sigma_y) \delta_{i,j-2} \right],
\label{bdi_model}
\end{align}
and the corresponding Bloch Hamiltonian is 
\begin{align}
H(k)=&\mqty( \kappa + \kappa' \sin(k) & t' e^{i2k} + t e^{ik}+m \nonumber\\t' e^{-i2k} + t e^{-ik}+m & -\kappa-\kappa' \sin(k)  )\nonumber \\
=& (t' \cos(2k)+t\cos(k)+m)\sigma_x \\
&+ (-t' \sin(2k)-t\sin(k))\sigma_y + (\kappa+\kappa'\sin(k)) \sigma_z.\nonumber
\end{align}

\begin{figure}[t]
	\centering
	\includegraphics[width=66mm]{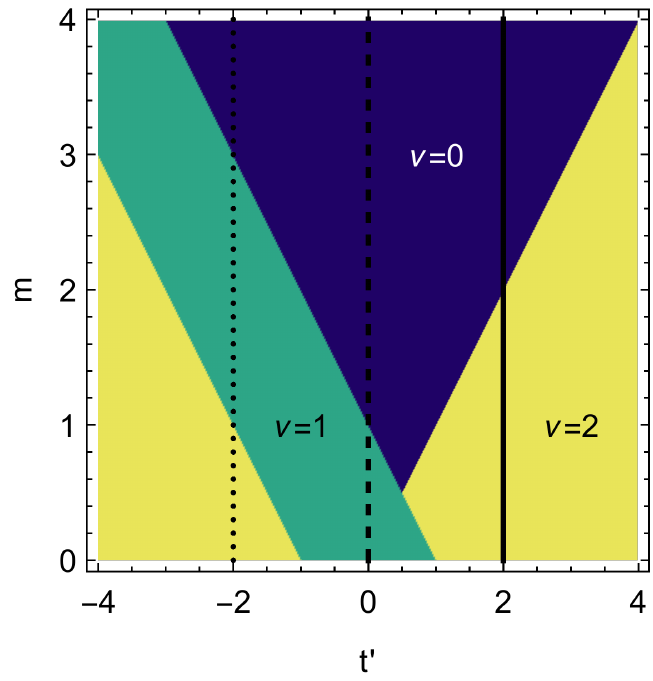}
	\caption{Phase diagram for the system in the $BDI$ class showing the winding number of each region --- plotted for parameters $t = 1,\kappa =\kappa'=0$. Three cuts of the phase diagram, which will be used in later figures,  are shown with continuous, dashed and dotted lines.}
\label{fig:bdi_phase_diagram}
\end{figure}

For $\kappa = \kappa' = 0$, it is in the BDI class, i.e. it has time-reversal ($T$), particle-hole ($C$), and chiral symmetry ($S$):
\begin{alignat}{2}
&T = \mathcal{K} ; \quad &&T H(-k) T^{-1} = H(k) \nonumber\\
&C = \sigma_z\mathcal{K} ; \quad &&C H(-k) C^{-1} = -H(k) \nonumber\\
&S = \sigma_z ; \quad &&S H(k)S^{-1} = -H(k) .
\end{alignat}
The corresponding phase diagram \cite{Song2014} is shown in Fig.~\ref{fig:bdi_phase_diagram} for $t=1$. 

\begin{figure}[h!]
\centering
\includegraphics[width=86mm]{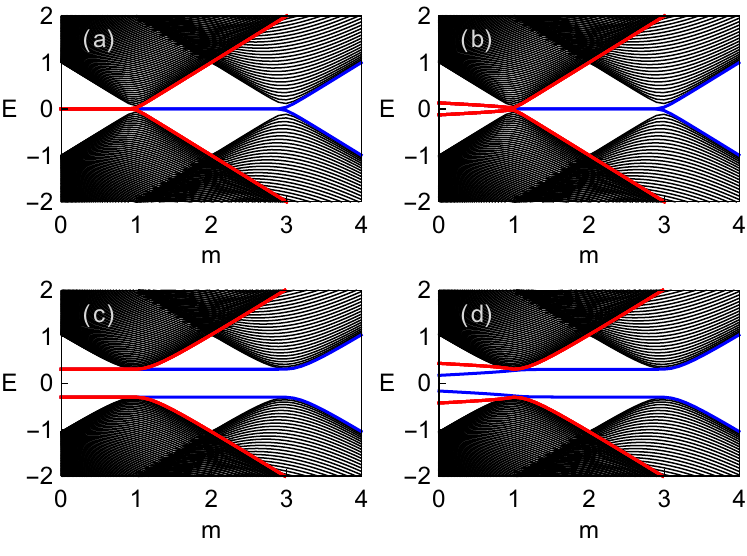}
\caption{Energy spectrum with open boundary conditions along the dotted path (i.e. $t=1,t'=-2$) of Fig.~\ref{fig:bdi_phase_diagram} plotted for different values of the symmetry-breaking terms (a) $\kappa =\kappa'=0$ , (b) $\kappa = 0, \kappa'=0.3$, (c) $\kappa = 0.3,\kappa'=0$ and (d) $\kappa = 0.3,\kappa'=0.3$}
\label{fig:zero_E_modes}
\end{figure}

The BDI class  in one spatial dimension  is characterized by a $\mathbb{Z}$ invariant, the winding number.
When open boundary conditions are imposed on the system, the number of symmetry-protected zero-energy edge modes is equal to the winding number. 
The behavior of the  edge modes  is  shown in Fig.~\ref{fig:zero_E_modes}, using the dotted path ($t=1,t'=-2$) marked in the phase diagram in figure~\ref{fig:bdi_phase_diagram}. 
Figure~\ref{fig:zero_E_modes}(a) shows the edge spectrum in the BDI class, with 4 (resp. 2) symmetry-protected zero modes for winding number 2 (1).

For $\kappa' \neq 0$, only particle-hole symmetry $C$ is preserved and the system belongs to symmetry class $D$. 
In one spatial dimension, the latter is characterized by a $\mathbb{Z}_2$ invariant.  
Consequently, adding such a term causes the zero modes in the $\nu=2$ phase to split pairwise and the phase becomes topologically trivial, indistinguishable from the phase with $\nu=0$.
This is shown in Fig.~\ref{fig:zero_E_modes}(b), where one clearly sees that the zero modes below $m=1$ are split. 

For $\kappa \neq 0$ only time-reversal $T$ is preserved and the system is in the AI class. 
In 1D this class is trivial and, thus,  all zero-energy modes split from zero energy. 
With both $\kappa \neq 0$ and $\kappa' \neq0$ the system has no local symmetry, thus belonging to the $A$ class.  
This symmetry class is also topologically trivial in one dimension. 
Fig.~\ref{fig:zero_E_modes}(c) and (d) show the edge spectrum for class AI and A, respectively. In both cases, the edge modes are split from zero for all values of $m$.

In Table \ref{fig:class_table} we summarize the different classes that this model can belong to for the different parameters, together with the correspondent symmetries and topological invariants in 1D.

\begin{table}[h]
\centering 
\begin{tabular}{|c |rrr| c|c|} 
\hline\hline 
Model &  $T$ & $C$ & $S$ & Class & inv.  \\  [1ex] 
\hline
$\kappa =0,\kappa'=0$ &  1 & 1 & 1 & BDI  & $\mathbb{Z}$\\  [1ex] 
$\kappa =0,\kappa' \neq0$ &  0 & 1 & 0 & D  & $\mathbb{Z}_2$ \\  [1ex] 
$\kappa \neq 0,\kappa'=0$ &  1 & 0 & 0 & AI & 0 \\  [1ex] 
$\kappa \neq 0,\kappa' \neq 0$ &  0 & 0 & 0 & A & 0  \\  [1ex] 
\hline \hline
\end{tabular}
\caption{Different Cartan classes for the model in Eq.~\eqref{bdi_model}, with the correspondent symmetries and topological invariants in 1D.} 
\label{fig:class_table}
\end{table}

\section{Geometric phases and bulk polarization}

The bulk polarization is a property which characterizes topological insulators in 1D. The bulk polarization is proportional to the surface charges of the system with open boundary conditions. In a topological insulator these are quantized because of the appearance of zero-energy edge states. Because of this, the bulk polarization is itself quantized, and therefore serves as a topological invariant.

In this section we review some relevant aspects of the bulk polarization and the modern theory of polarization, which gives a geometrical description of the polarization \cite{Resta1992,KingSmith1993,Vanderbilt1993,Resta1997}. 

Consider a generic, quadratic Hamiltonian in one dimension
\begin{equation}\label{eq:quadr_Ham}
\mathcal{H} = \sum_{ij,\alpha\beta} c_{i\alpha}^\dagger H_{ij,\alpha \beta}c_{j\beta}.
\end{equation}
We obtain its single-particle eigenstates by
\begin{align}
\sum_{j\beta}H_{ij,\alpha\beta} \psi_{p\mu}^{j\beta} = E_{p\mu} \psi^{i\alpha}_{p\mu},
\end{align}
where $[U]_{j\beta,p\mu} = \psi_{p\mu}^{j\beta}$ is the unitary matrix that diagonalizes $H$. In translational invariant systems $\psi_{k\mu}^{j\beta} = e^{ikj}u_{k \mu}^{\beta}$, where $u_{k \mu}^{\beta}$ are the components of the eigenstates, $\ket{u_{k\mu}}$, of the Bloch Hamiltonian and $k$ is the momentum. For simplicity we consider a unit cell with a single site. One can now compute the Zak phase \cite{Zak1989}.
 For a two-band  model, it is defined as the geometric phase acquired by the occupied state $\ket{u_k}$ as it winds around the Brillouin zone,
\begin{equation}
\gamma = \int_{0}^{2\pi} {\rm d}k\, i A_k, 
\label{eq:zak_phase}
\end{equation}
where $A_k = \bra{u_k}\partial_k \ket{u_k}$ is the Berry connection \cite{Berry1984}. 
The generalization to multi-band models is straightforward. 

The Zak phase is only gauge invariant modulo $2\pi$. However, for two different states defined by a parameter $\lambda$ --- assuming $A_k(\lambda)$ is smooth in the path connecting them --- the change in the Zak phase can be computed as
\begin{equation}
\Delta {\gamma_{\lambda_i \lambda_f}} = \int_{\lambda_i}^{\lambda_f} d\lambda\int_{0}^{2\pi} dk \, \Omega_{\lambda k},
\label{eq:change_zak_phase}
\end{equation}
where 
\begin{align}\label{eq:BerryCurvature}
\Omega_{\lambda k} = \partial_\lambda A_k(\lambda,k) - \partial_k A_\lambda(\lambda,k)
\end{align}
 is the Berry curvature. The Berry curvature is fully gauge invariant and, therefore, the change in the Zak phase in Eq.~\eqref{eq:change_zak_phase} is defined in $\mathbb{R}$. 
The Zak phase itself, as defined in Eq.~\eqref{eq:zak_phase}, can be shown to be proportional to the winding number for a certain gauge (see Appendix \ref{appendix:winding_ssh}), which means that it carries physical information beyond modulo $2\pi$. Note that other ways of computing the Zak phase rely on computing $e^{i\gamma}$ instead, such that the result is always defined only modulo $2\pi$.

The derivative of the polarization with respect to $\lambda$ was obtained in Ref.~\cite{KingSmith1993} as
\begin{equation}\label{eq:Bloch_polarization}
\partial_\lambda \mathcal{P}^{\rm Bloch} = \int_{0}^{2\pi} \frac{dk}{2\pi} \, \Omega_{\lambda k},
\end{equation}
which allows us to make the identification $ \mathcal{P}^{\rm Bloch} = \gamma/2\pi$. The polarization itself is however not an observable. The relevant quantities are the derivatives or changes in the polarization with respect to external parameters. These result in currents and charge transport, which are the only measurable quantities. Similar to how only difference in the Zak phases between two states can be measured \cite{Atala2013}. 

Considering a system with periodic boundary conditions, one can also compute the geometric phase obtained by threading a $U(1)$ flux through the ring. This allows us to define a polarization in the absence of translational invariance. Depending on how the flux is introduced we obtain different polarizations with different physical meanings. These different polarizations, their relations and physical interpretations were studied in a recent article by Watanabe and Oshikawa \cite{Watanabe2018}. We focus on two polarizations, $\mathcal{P}$ and $\tilde{\mathcal{P}}$, obtained by introducing the flux in two different ways.

If the flux is introduced homogeneously through a vector potential $A_x = \Phi/L$ we can define the polarization
\begin{equation}\label{polarization_inhom}
\mathcal{P} = \int_0^{2\pi}\frac{d\Phi}{2\pi} i\bra{\Psi^\Phi}\partial_\Phi \ket{\Psi^\Phi} + \frac{1}{2\pi}\Im\ln \bra{\Psi^0}e^{2\pi i \hat{P}} \ket{\Psi^{2\pi}}, 
\end{equation}
where $\ket{\Psi^{\Phi}}$ is the ground state in the presence of flux and $\hat{P}=\frac{1}{L}\sum_{j\alpha} j\hat{n}_{j\alpha}$ is the polarization operator, with $\hat{n}_{j\alpha}$ being the number operator. The second term is needed to make the expression gauge invariant. The derivative of this polarization gives the average current along the chain.

This is equivalent to the polarization obtained by Resta \cite{Resta1997} as
\begin{align}
\mathcal{P} = \frac{1}{2\pi} \Im \,\ln\bra{\Psi^{\Phi=0}}e^{2\pi i \hat{P}}\ket{\Psi^{\Phi=0}}.
\label{eq:p_resta}
\end{align}
Expression \eqref{eq:p_resta} is particularly useful because it can be easily expressed in terms of single-particle eigenstates as
\begin{align}
\mathcal{P} = \frac{1}{2\pi}\Im \ln \, {\rm det }\!' \, S,
\label{eq:polarization_resta}
\end{align}
where the matrix $S$ is given by
\begin{equation}
S_{p\mu,q\nu} = \sum_{j\alpha} \psi_{p\mu}^{j \alpha \, \ast} e^{i\frac{2\pi}{L}j}\psi_{q\nu}^{j \alpha},
\end{equation}
and  ${\rm det }\!' $ indicates that the determinant is restricted to the space of \emph{occupied} single-particle states. Below we will compare the bulk polarization obtained using our method to the one obtained using equation~\eqref{eq:polarization_resta}.

If the flux is introduced via twisted boundary conditions at the seam, i.e. in the bond between sites $j=1$ and $j=L$, this is equivalent to performing the gauge transformation
\begin{align}\label{eq:Psi_tilde}
\ket{\tilde{\Psi}^\Phi}=e^{i\Phi \hat{P}}\ket{\Psi^\Phi},
\end{align}
which makes $\ket{\tilde{\Psi}^\Phi}$ fully periodic in $\Phi$. We can now define the bulk polarization
\begin{equation}
\tilde{\mathcal{P}} = \int_0^{2\pi}\frac{d\Phi}{2\pi} i\bra{\tilde{\Psi}^\Phi}\partial_\Phi \ket{\tilde{\Psi}^\Phi},
\label{eq:p_tilde}
\end{equation}
whose derivative gives the current flowing through the seam \cite{Watanabe2018}. This bulk polarization is related to the previous one by 
\begin{equation}
\mathcal{P}= \tilde{\mathcal{P}}+\bar{\mathcal{P}}_0,
\end{equation} 
where
\begin{equation}
\bar {\mathcal{P}}_0= \int_0^{2\pi}\frac{d\Phi}{2\pi} \bra{\tilde{\Psi}^\Phi}\hat{P}\ket{\tilde{\Psi}^\Phi}.
\end{equation} 
Similar to the discussion about the Bloch polarization below Eq.~\eqref{eq:Bloch_polarization}, both $\mathcal{P}$ and $\tilde{\mathcal{P}}$ are defined modulo 1, while their changes are defined in $\mathbb{R}$. For translationally invariant systems one finds that
\begin{equation}
\mathcal{P} = \mathcal{P}^{\rm Bloch}+\frac{L-1}{2}\nu \, {\rm mod} \, 1
\end{equation}
where here $\nu$ denotes the number of occupied bands \cite{Watanabe2018}.

\section{Entanglement spectrum}
\label{section:ES}

\begin{figure}[t]
	\centering
	\includegraphics[width=85mm]{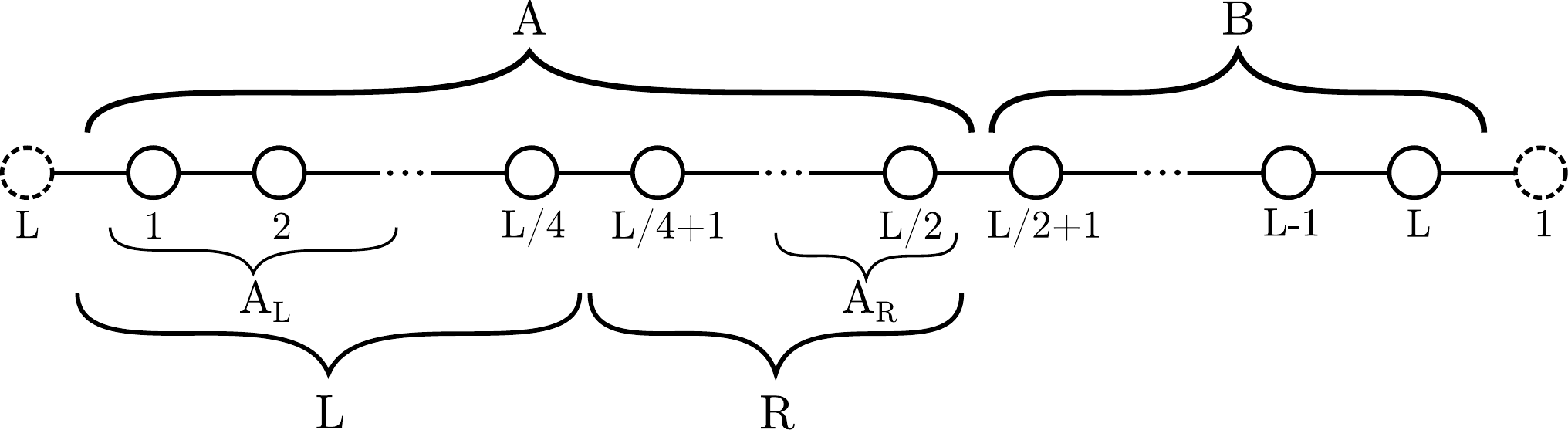}
	\caption{Schematic of the chain with periodic boundary conditions considered, where the different sites are labeled by their position. The bipartition into regions $A$ and $B$ is shown. Other subregions of $A$ considered in the text are also shown. }
\label{fig:chain}
\end{figure}

Let us now proceed to discuss the ES. 
An important quantity to consider is the correlation matrix, which in position space is defined by the ground state expectation value
\begin{align}
C_{ij}^{\alpha \beta} = \expval{c_{i\alpha}^\dagger c_{j\beta}}.
\end{align}
For a system where all single-particle eigenstates with energies $E<0$ are occupied, the correlation matrix can be written (see  Appendix~\ref{app:CM}) in terms of the Hamiltonian as
\begin{align}\label{eq:corr_mat2}
C_{ij}^{\alpha \beta} =& \frac{1}{2}\left[I - H/ (H^2)^{1/2} \right]_{ij, \alpha \beta}.
\end{align}

In this paper, we are mainly interested in the spectrum of the correlation matrix when restricted to a spatial sub-region $A$ (its complement will be denoted by $B$ in the following), see figure~\ref{fig:chain}. When restricted to sub-region A, the boundaries of A act as virtual boundaries to the system.
Following Ref.s~\cite{Huang2012,Huang2012-2}, we refer to this spectrum as the entanglement occupancy spectrum (EOS).
We denote the eigenvalues of the EOS by $\xi_j$. 
Topological phases are characterized by (symmetry-protected) zero-energy modes in the edge spectrum and $\xi=\frac 1 2$ virtual edge modes in the EOS.

Following Ref.~\cite{Peschel2003,Peschel2008} , one can write the  reduced density matrix as 
\begin{align}\label{eq:red_dens_mat}
\rho_A&=\mathcal{K} \exp(-\mathcal H),
\end{align}
where $\mathcal{K}$ is a normalization constant and $\mathcal{H}$ is a quadratic Hamiltonian, referred to as entanglement Hamiltonian. 
The spectrum of the reduced density matrix is the ES. The single-particle eigenvalues $\epsilon_j$ of $\mathcal{H}$, referred to as entanglement energies, form the single-particle ES. They are related to those of the EOS by 
\begin{align}\label{eq:xi_eps}
\xi_j &=\left(e^{\epsilon_j}+1\right)^{-1}, 
\end{align}
the corresponding eigenstates are the same. Consequently, the EOS is in one-to-one correspondence to the single-particle ES \cite{Fidkowski2010entanglement}.
Eq.~\eqref{eq:xi_eps} implies that, for non-interacting systems, the full information of the ES is contained in the spectrum of the subsystem correlation matrix. 
The latter is much simpler to interpret. 
Consequently, we will focus on the EOS in the remainder of the paper and only mention the ES when necessary. 

One property of the subsystem correlation matrix we will make use of later is that, for sufficiently large systems, its eigenstates are found to be exponentially localized on either virtual edge if $\xi$ is away from $0$ and $1$, or they are found to be bulk modes if the correspondent eigenvalues are exponentially close to $\xi = 0,1$ \cite{Peschel2008}.

A few comments are needed regarding our choice of  bipartition. 
In the following sections, we show how the Zak phase can be recovered from the EOS.
We also provide a simple formula in terms of the EOS eigenvalues that is identical to the Zak phase in the thermodynamic limit.  
However, for a generic, finite-size system, there will be finite-size discrepancies between our formula and the Zak phase --- for a finite-size scaling analysis see figure \ref{huang}(d). 
In order to reduce these, we are going to choose subsystem $A$ as half the system in the remainder of the manuscript.
The results (in the thermodynamic limit) do not depend on this choice. 
\section{Bulk Polarization in the EOS}\label{sec:bulkpolarization}

Let us first review previous results on the relation between the EOS and the bulk polarization (or Zak phase). 
For systems in 1D protected by chiral or particle-hole symmetries the Zak phase is a topological invariant. It is zero whenever there is an even number  of eigenvalues per virtual edge at $\xi = 1/2$, and $\pi$ when the number is odd~\cite{Asboth2016}. 
In the absence of symmetry-protection, much less is known. 
Ryu and Hatsugai considered the fully dimerized SSH chain with broken chiral symmetry~\cite{Ryu2006}.  
This model is rather special in that there is only a single pair of eigenvalues in the EOS that is not strictly identical to $0$ or $1$: these are $\xi$ and $1-\xi$, related due to translational symmetry. 
The authors could show that the value of one of these `midgap states' is identical to the Zak phase divided by $2\pi$ --- or alternatively $\tilde{\mathcal{P}}^{\rm Bloch}$ --- although they do not specify to which of the two eigenvalues it corresponds. 
Their result is a particular limit of equation~\eqref{eq:chi_integrable}, which we discuss in the next section. 

In Ref.s~\cite{Huang2012,Huang2012-2} the authors considered a two-dimensional Chern insulator and noted that there was a similarity between the Zak phase and the virtual topological edge states that connect the EOS values at 0 to those at 1. 
Interpreting this as a one-dimensional system with a parameter, we can explain this behavior by Eq.~\eqref{eq:chi_left}, noting that there is a single pair of eigenvalues that dominates the sum. 
Note that the observation of Ref.s~\cite{Huang2012,Huang2012-2} is  particular to systems with Chern number 0 or $\pm 1$, i.e. where there are only few midgap states in the EOS. 
It fails for systems with higher Chern numbers,  for which there are several terms in Eq.~\eqref{eq:chi_left} with comparably large contributions and, consequently, the Zak phase deviates considerably from the virtual topological edge states.

\subsection{Systems with equispaced entanglement energies}

We first discuss systems for which the entanglement energies of \emph{either} virtual edge are equispaced. That is, they are given in the thermodynamic limit by $\varepsilon_{n\alpha} = \varepsilon_{0\alpha}+n\delta_\alpha$, where $\alpha = L,R$ labels the two virtual edges. 
This is a feature of integrable systems \cite{Peschel1999},  e.g. nearest neighbor hopping models such as the SSH chain, and it applies to the Hamiltonian in Eq.~\eqref{bdi_model} when $t'=0$. 
In this case, one can obtain the Zak phase in a very simple fashion: We reorder the eigenvalues of the EOS by magnitude, $\xi_1 < \xi_2 < ...< \xi_{L_AM}$, and compute
\begin{equation}
\chi = \sum_{j=1}^{L_A M} \xi_{2j-1} \, {\rm mod} \, 1,
\label{eq:chi_integrable}
\end{equation}
where $L_A$ is the length of subsystem $A$ and $M$ is the number of orbitals per site. In the thermodynamic limit, $\chi$ becomes identical to either $\tilde{\mathcal{P}}^{\rm Bloch}$ or $1-\tilde{\mathcal{P}}^{\rm Bloch}$, 
\begin{equation}
\lim_{L \rightarrow \infty} \abs{\tilde{\mathcal{P}}^{\rm Bloch}} = \lim_{L \rightarrow \infty} \chi \, {\rm mod} \, 1.
\end{equation}

In the special case where there is only one eigenvalue per edge that is $\neq 0,1$ in  the EOS, like in the fully dimerized limit of the SSH chain, we recover the equality between one of the two midgap eigenvalues and the Zak phase found in reference ~\cite{Ryu2006}. 

The issue concerning the sign ambiguity will be discussed and resolved in the next section. 
The origin of this ambiguity can be traced back to certain properties of the EOS. This sign is irrelevant in the case of symmetry-protected states where the Zak phase is quantized to $0$ or $\pi$.

\begin{figure}[t]
\centering
\includegraphics[width=86mm]{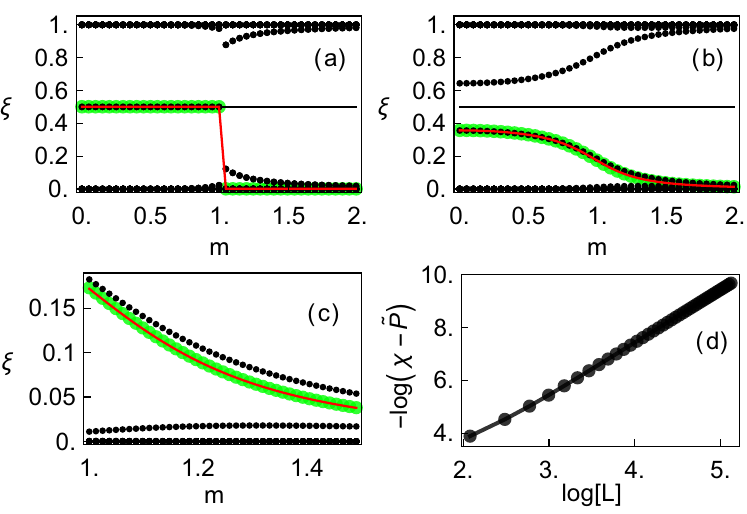}
\caption{In (a) and (b)  we show the EOS (black), $\chi$ (green) and $\tilde{\mathcal{P}}^{\rm Bloch}$ (red) for a cut in the phase diagram through the $\nu = 1 \rightarrow \nu = 0$ transition (dashed line in Fig.~\ref{fig:bdi_phase_diagram}, for $t = 1,t'=0$), for parameters  (a) $\kappa =\kappa'=0$ and (b) $\kappa = 0.3, \kappa'=0$. Computed for $L=40$. In (c) we zoom in on a region of the plot in (b) to show the splitting between $\tilde{\mathcal{P}}^{\rm Bloch}$ and the eigenvalue closest to $\xi=1/2$. In (d) we show a scaling plot of the difference between $\chi$ and $\tilde{\mathcal{P}}^{\rm Bloch}$ for parameters $t=1,t'=0,m=1$, $\kappa = 0.3, \kappa'=0$ and increasing system size to show that the difference vanishes in the thermodynamic limit. }
\label{huang}
\end{figure}

In Fig.~\ref{huang} we show  the EOS along the dashed cut in the phase diagram of Fig.~\ref{fig:bdi_phase_diagram}, for $t'=0$. 
Along this line, our model is equivalent to an SSH chain. 
In the presence of chiral and translation symmetry, all eigenvalues of the EOS are doubly degenerate: one eigenvalue from each virtual edge, such that $\chi$ from Eq.\eqref{eq:chi_integrable} will pick up one eigenvalue from each pair. Because of chiral symmetry, the sum of the two eigenvalues related by the symmetry gives always $1$ so they do not contribute to $\chi$. The only contribution to $\chi$ is the one from the eigenstates with $\xi=1/2$ that do not have a chiral partner, see figure~\ref{huang}(a). Here we see explicitly the known relation between the Zak phase and the number of $\xi = 1/2$ modes.
Note that the full sum  in Eq.~\eqref{eq:chi_integrable}  evaluates to zero in the trivial regime despite the presence of modes with entanglement occupancy $0<\xi<1$. 

When breaking chiral symmetry, as shown in figure~\ref{huang}(b), the EOS will no longer be doubly degenerate. Note that, even with the broken symmetry, the bulk polarization shows plateaus that point to a possible nearby topological transition.
For small $m$, there are two dominant midgap eigenvalues, and the bulk polarization follows the lower one very closely.
When this midgap eigenvalue approaches  0, the contribution from the other modes becomes more relevant and the bulk polarization starts to deviate from the dominant eigenvalue, see figure~\ref{huang}(c). In Fig.~\ref{huang}(d) we show how the difference between $\chi$ and the bulk polarization decreases with system size, showing that they are equal in the thermodynamic limit. Note that their discrepancy is very small already for small system sizes. 

\subsection{General case}

For general non-interacting gapped systems we find (see Appendix \ref{app:pollmann}) that it is only the eigenstates that localize on the left virtual edge ($A_L$ in Fig.\ref{fig:chain}) that should contribute to $\chi$. The inclusion of bulk modes is irrelevant since they have $\xi = 0,1$. We can then compute
\begin{equation}
\chi = \sum_{\mu \in L} \xi_\mu \, {\rm mod} \, 1. 
\label{eq:chi_left}
\end{equation}
the sum is performed over the subspace $L$ formed by the eigenstates with $\expval{\hat{x}}<L/4$, where $L/4$ is the middle point of region $A$. It does not matter what threshold we use as long as we include all left-edge states and exclude all right-edge states. In practice only a few eigenvalues give a significant contribution to $\chi$.

In the thermodynamic limit we obtain
\begin{equation}
\lim_{L \rightarrow \infty} \tilde{\mathcal{P}} = \lim_{L \rightarrow \infty} \chi \, {\rm mod } \, 1.
\label{eq:ptilde_eq_chi}
\end{equation}
Note that in order to compute $\chi$ correctly the eigenstates with $\xi \neq 0,1$ need to be localized in order for us to select the correct eigenvalues. It might be --- like in Fig.\ref{huang}(a) --- that the spectrum has degenerate pairs of left and right eigenstates and the numerical diagonalization mixes them. This case is trivial, since we know that there is one left and one right eigenstate, but if the degeneracy is four-fold or higher we must obtain the localized eigenstates. Localizing the bulk eigenstates that are exponentially close to $\xi = 0,1$ is more difficult but not doing so only results in an exponentially small error. 

Before proceeding, lets review again the result so far. First note that for translation invariant systems, the ES of the right and the left virtual edge are related by an overall '-' sign, i.e. for each eigenvalue $\epsilon$ on the right virtual edge, there is a corresponding value $-\varepsilon$ on the left one. 
Since the spectra are equispaced, this implies that the two virtual edge spectra are merely shifted with respect to each other.
If the shift is zero, e.g.  in presence of a symmetry, summing every other eigenvalue is equivalent, modulo 1, to summing all the left eigenvalues. For a finite shift, summing all odd eigenvalues corresponds to either summing all left eigenvalues or all right eigenvalues. In general, this implies that $\chi$, as defined in Eq.~\eqref{eq:chi_integrable}, only gives the Zak phase up to an overall sign. This issue can be resolved by determining the localization of a single eigenstate.

\begin{figure}[t]
\centering
\includegraphics[width=86mm]{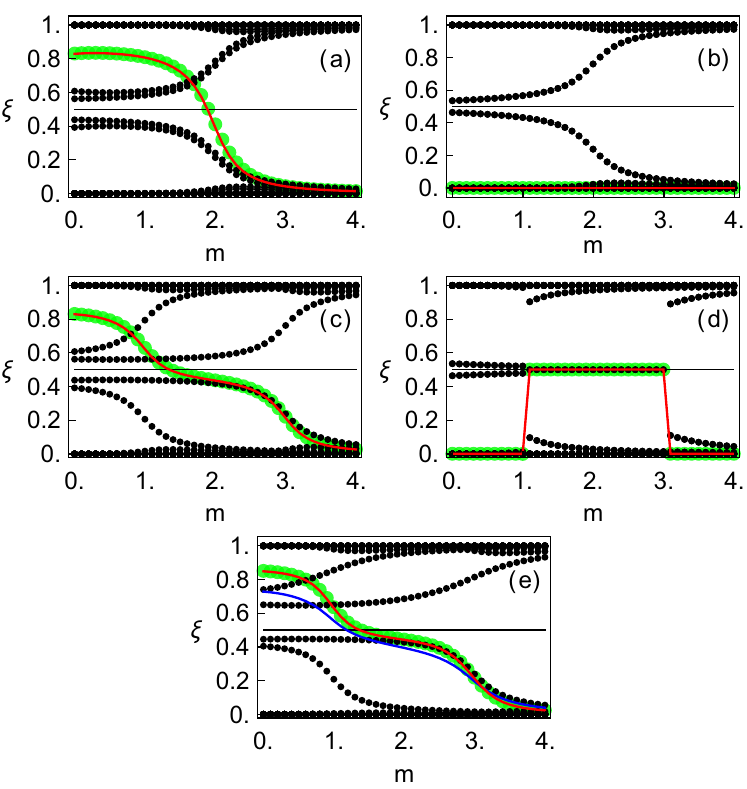}
\caption{We show the EOS (black), $\chi$ (green) and $\tilde{\mathcal{P}}$ (red) computed for $L=40$ sites. (a) and (b) for the continuous line in the phase diagram, $t=1,t'=2$  with the symmetry-breaking parameters (a) $\kappa=0.3,\kappa'=0$ and (b) $\kappa=0,\kappa'=0.3$. (c) and (d) for the dotted line in the phase diagram, $t=1,t'=-2$  with the symmetry-breaking parameters (c) $\kappa=0.3,\kappa'=0$ and (d) $\kappa=0,\kappa'=0.3$. (e) is the same as (c) but with a position dependent $\kappa_i=(i+L/4 \, {\rm mod}\, L)/L$. We also plot $\mathcal{P}+1/2$ (blue) to showcase that $\chi$ is indeed equal to $\tilde{\mathcal{P}}$ and not $\mathcal{P}$. }
	\label{2}
\end{figure}

In Fig.~\ref{2} we show two examples that highlight the importance of the localization structure of the EOS. 
In figures~\ref{2}(a) and (b), we compute the EOS for varying $m$ at  $t=1$, $t'=2$, indicated by the solid line  in the phase-diagram in Fig.~\ref{fig:bdi_phase_diagram}, in the presence of two different symmetry breaking terms: (a) $\kappa =0.3$, $\kappa'=0$ and (b) $\kappa=0$, $\kappa'=0.3$.  
For $\kappa=\kappa'=0$, there is a phase transition $\nu = 2 \rightarrow 0 $ at $m=2$, whereas both choices of symmetry breaking render the system trivial  for all $m$.
However, the two symmetry breaking terms split the four-fold degenerate $\xi=1/2$ midgap states very differently, thus resulting in completely different Zak phases despite the superficial resemblance of their respective EOS.  
While $\kappa'$ splits the four-fold degenerate states into two two-fold degenerate pairs at opposite edges whose contribution cancels (modulo 1), $\kappa$ split them into two $left-left$ and $right-right$ pairs, with a finite contribution to the Zak phase. 
In figures~\ref{2}(c) and (d), we consider the same symmetry breaking terms, but now for $t=1$ and $t'=-2$, where the BDI systems shows a phase transition  $\nu=2\rightarrow 1\rightarrow 0$. 
Also here, Eq.~\eqref{eq:chi_left} correctly reproduces the Zak phase for all values of $m$. 
Note that in Fig.s~\ref{2}(a) and (b) the observation made in Ref. \cite{Huang2012}, i.e. that the there is a similarity between the Zak phase and the EOS, does not apply anymore. 

Since $\chi$ is defined in position space our results are also valid for systems without translationally invariance. We show this in Fig.~\ref{2}(e) where we show the EOS, $\chi$, $\mathcal{P}$ and $\tilde{\mathcal{P}}$ along the dotted cut in the phase diagram with an added position dependent symmetry breaking term $\kappa_i = (i+L/4 \, {\rm mod}\, L)/L$. We show that Eq.~\eqref{eq:ptilde_eq_chi} holds without translational invariance, where $\mathcal{P}$ now differs from $\chi$ and $\tilde{\mathcal{P}}$.

This relation between the Zak phase and the EOS is a consequence of an identity between the Zak phase and the many-body ES previously found for infinite chains \cite{Zaletel2014}. In Appendix A we modified this derivation to account for the periodic boundary conditions we use and show how their result simplifies greatly when expressing it in terms of the EOS, resulting in equation~\eqref{eq:ptilde_eq_chi}.


\section{Alternative Bulk Polarization and Chern Numbers}

As mentioned above, the bulk polarization is only defined modulo 1. This is obvious when it is defined in terms of a geometric phase, which is only defined modulo $2\pi$. In our formulation in Eq.~\eqref{eq:ptilde_eq_chi}, this is reflected in an ambiguity in the number of bulk modes with $\xi = 1$ that are included in $\chi$. Being defined only modulo $1$, the bulk polarization cannot differentiate the phases $\nu=0$ and $\nu=2$, even though the EOS of both phases is distinct. Another issue due to it being defined in $\mathbb{R}$ mod $1$ appears when we try to follow its evolution over a path $C_\lambda$. Between any two points $\lambda$ and $\lambda'$ in this path it is not possible to tell if the bulk polarization has increased or decreased by only looking at $\tilde{\mathcal{P}}(\lambda)$ at these specific points.
 This latter drawback becomes relevant when computing Chern numbers.

In an attempt to solve these issues we introduce an alternative bulk polarization given by
\begin{equation}\label{eq:tpo}
\tpo = \int_0^{2\pi}\frac{d\Phi}{2\pi} i\brapsio{\Phi}\partial_\Phi \ketpsio{\Phi}.
\end{equation}
where $\ketpsio{{\Phi=0}}$ is the ground state when the (previously periodic) chain is opened between sites $j=L/2$ and $j=L/2+1$ and 
\begin{equation}
\ketpsio{\Phi} =e^{-i \Phi \hat{N}_A} \ketpsio{0}.
\label{eq:gauge}
\end{equation}
When introducing the flux in the system with periodic boundary conditions (See Eq.~\eqref{eq:Ground_state_flux}) there is an ambiguity due to the bulk modes. 
This ambiguity disappears when we introduce the flux as above in the open chain --- we simply include all the bulk modes. 
Note that $\Phi$ is not a magnetic flux but should be treated as an additional parameter, and therefore $\tpo$ is not the polarization of the open chain. $\tpo$ can be obtained in an even simpler way than $\tilde{\mathcal{P}}$ as 
\begin{align}\label{eq:phi_independence}
\tpo =&  \int_0^{2\pi}\frac{d\Phi}{2\pi} \brapsio{0} \hat{N}_A \ketpsio{0} \nonumber \\
=&\sum_{\alpha,j\in A}\brapsio{0} c_{j\alpha}^\dagger c_{j\alpha} \ketpsio{0} \nonumber\\
=&\sum_{\alpha,j\in A} C_{\rm o,j\alpha,j\alpha}\nonumber \\
=& {\rm Tr}[C_{A,\rm o}],
\end{align}
where we used Eq.~\eqref{eq:gauge} and ($C_{A,\rm o}$)$C_{\rm o}$ denotes the (subsystem) correlation matrix of the open chain.  

The idea is that in the process of opening the chain, which can be done adiabatically, all eigenvalues that are related to the right virtual edge ($A_R$ in Fig.~\ref{fig:chain}) are pushed to $\xi=0,1$. 
As a result, the sum of all eigenvalues of $C_{A,o}$ (modulo 1) is equal to $\chi$ \eqref{eq:chi_left}. \footnote{ This is strictly speaking only true in the thermodynamic limit, where we avoid the finite size effects due to opening the chain.} Therefore, in the thermodynamic limit, we have
\begin{equation}
\lim_{L \rightarrow \infty }\tilde{\mathcal{P}} = \lim_{L \rightarrow \infty } \tpo \, {\rm mod} \, 1.
\end{equation}

The advantage of using $\tpo$ is that the contribution from the 'bulk' modes is constant as long as no states cross the Fermi energy, i.e. the system remains gapped.  
The quantity $\chi$ \eqref{eq:chi_left}, on the other hand, is only independent of the bulk mode contribution when the modulo 1 is included. 
The consequence is that for paths $C_\lambda$ in parameter space, for which the energy spectrum is gapped, $\tpo (\lambda)$ is defined in $\mathbb{R}$, such that the total change in $\tpo$ along this path can be obtained knowing $\tpo$ only in the initial and final states, $\Delta \tpo ({C_\lambda}) = \tpo(\lambda_f) - \tpo(\lambda_i)$. Note, however, that $\tpo$ is gauge-invariant only when taken modulo 1. This is similar to the case of the Zak phase, which is known to be equal to the winding number for certain gauges (see Appendix~\ref{appendix:winding_ssh}). It encodes more information than what is accessible when considered modulo $2\pi$.

This method resolves most of the issues found in section \ref{sec:bulkpolarization}, but unfortunately it cannot be applied when the system is in a non-trivial topological phase. When we open the chain for a system in a topological non-trivial state, zero-energy modes will appear. The ground state of the closed chain has several equivalent degenerate ground states for the open chain, so that $C_{A,\rm o}$ is not well-defined at these points. 
One might think that this problem can be circumvented by introducing symmetry-breaking term that render the system trivial and extrapolating $\tpo$ towards the topological phase. 
However, this extrapolation is again not necessarily unique, as there might be integer jumps in $\tpo$ when the edge-modes cross zero-energy. 
As we will see below these jumps in $\tpo$ are branch cuts that provide the correct description of the system.

\subsection{SSH chain}

\begin{figure}[t]
\includegraphics[width=86mm]{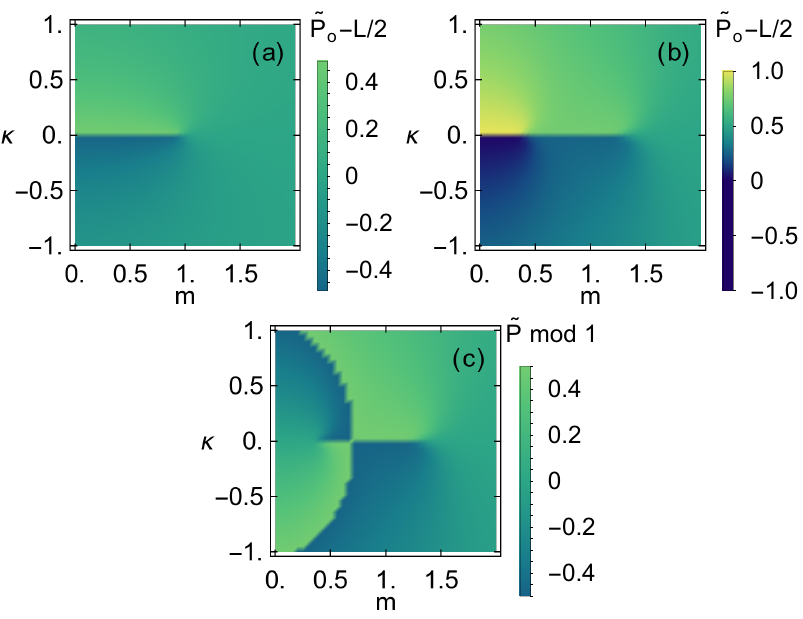}
\caption{(a) and (b) show the bulk polarization $\tpo-L/2$ computed for $t=1,\kappa'=0$ and $L=40$ sites, with (a) $t'=0$ and (b) $t'=-2$. $\tpo$ exhibits branch cuts that indicate the presence of zero-energy modes. $\tpo$ is able to differentiate between the phases $\nu=2$ and $\nu=0$. In (c) we show $\tilde{\mathcal{P}}$ mod $1$ for the same parameters of (b) to illustrate the advantage of using $\tpo$. }
	\label{ssh_chern}
\end{figure}

Consider first the simple case of the SSH chain. When we include the chiral-symmetry breaking parameter $\kappa$ the Hamiltonian has the symmetry
\begin{equation}\label{eq:sym}
SH(\kappa)S^\dagger = -H(-\kappa), \quad S C_A(\kappa) S^\dagger = \mathbb{I}-C_A(-\kappa),
\end{equation}
so $\tpo(-\kappa) = L-\tpo(\kappa)$. Unless $\tpo(0) = L/2$, which is the result expected for a trivial insulator, the result will be discontinuous at $\kappa = 0$. We can see this in Fig.~\ref{ssh_chern}(a) where we show $\tpo$ in the parameter space $(m,\kappa)$. The presence of the zero-mode for $m<1$ appears in $\tpo$ as a branch cut. The appearance of a branch cut in the $(m,\kappa)$ parameter space is natural but its position depends on the choice of gauge (see Appendix~\ref{appendix:winding_ssh}).

The relation between the zero-modes and discontinuities in $\Tr[C_{A,o}]$ was already observed in the context of Chern insulators \cite{Alexandrinata2011}. The appearance of the branch cut is the result of the degenerate point at $(m=1,\kappa=0)$ having an associated Chern number $C=1$ \cite{Asboth2016}.

For simplicity, assume that the branch cuts appear along $\kappa=0$, such as depicted in Fig.s~\ref{ssh_chern} (a) or (b). The generalization to more complicated scenarios is straightforward. 
Consider a counterclockwise loop around the degenerate point at $(m=t,\kappa=0)$ starting from the branch cut parametrized by $\theta$,
\begin{align*}
& m(\theta) = 1+\frac{1}{2}\sin(\theta-\frac{\pi}{2}) \\
& \kappa(\theta) = -\sin(\theta).
\end{align*}
The Chern number is then defined as
\begin{equation}
C = \frac{1}{2\pi}\oint d\theta\oint d\Phi \, \Omega_{\theta\Phi},
\label{eq:chern}
\end{equation} 
with the Berry curvature $\Omega_{\theta\Phi}$ defined in equation~\eqref{eq:BerryCurvature}. 
From  Eq.~\eqref{eq:tpo}, it follows that the Berry connection for the flux is simply $A_\Phi(\theta,\Phi) = \tpo(\theta)$, so we now need to compute $A_\theta(\theta,\Phi)$. 

We can expand $\ket{\tilde{\Psi}^{\Phi = 0}_o}$ in the eigenbasis of the number operator $\hat{N}_A$, denoted by $\{\ket{j}\}$, as
\begin{equation}
\ket{\tilde{\Psi}^{\Phi = 0}_o(\theta)} = \sum_j c_j(\theta)\ket{j},
\end{equation}
where $c_j(\theta) = \Big\langle j\ket{\tilde{\Psi}^{\Phi = 0}_o(\theta)}$. The flux can then be introduced as
\begin{equation}
\ket{\tilde{\Psi}^{\Phi }_o(\theta)} = \sum_j c_j(\theta)e^{-i \Phi N_A^j}\ket{j}.
\end{equation}
Computing the derivative of the ground state with respect to the parameter $\theta$ gives
\begin{align}
\partial_\theta \ketpsio{\Phi(\theta)} =& \sum_j [\partial_\theta c_j (\theta)]e^{-i\Phi N_A^j}\ket{j}.
\end{align}
The correspondent Berry connection then gives
\begin{align}
\brapsio{\Phi(\theta)}\partial_\theta \ketpsio{\Phi(\theta)} =& \sum_{lj} c^*_l(\theta) [\partial_\theta c_j(\theta)] \bra{l} e^{i\Phi (N_A^l-N_A^j)}\ket{j}\nonumber\\
=& \sum_j  c^*_j(\theta) \partial_\theta c_j(\theta) \nonumber\\
=&\brapsio{0(\theta)}\partial_\theta \ketpsio{0(\theta)}, 
\end{align}
i.e. it is independent of $\Phi$. 

The integral in Eq.\eqref{eq:chern} is performed on the surface of the torus defined by $(\theta,\Phi)$. Because of the branch cut, the Berry connection $A_\Phi(\theta)$ cannot be made smooth for the whole torus and therefore we cannot directly apply Stokes theorem. Instead, we split the torus in two cylinders, $S_1$ with $\theta\in[\theta_0,2\pi-\theta_0]$ and  $S_2$ with $\theta\in[2\pi-\theta_0,2\pi+\theta_0]$ and we change the gauge in $\boldsymbol{A}(\theta)$ such that it is smooth in each cylinder. We can now apply Stokes theorem to the two cylinders independently \cite{Kohmoto1985}. For $S_1$ we have
\begin{align}
C_1 =& \frac{1}{2\pi} \int_{S_1} d\boldsymbol{S}\cdot (\nabla \times \boldsymbol{A}) \nonumber\\
=&\frac{1}{2\pi} \int_{\partial S_1} d\boldsymbol{l}\cdot \boldsymbol{A}(\theta,\Phi) \nonumber\\
=& \frac{1}{2\pi} \int_{\theta_0}^{2\pi-{\theta_0}} d\theta A_\theta(\theta,0) +\frac{1}{2\pi}\int_0^{2\pi} \frac{d\Phi}{2\pi} A_\Phi (2\pi-{\theta_0},\Phi) \nonumber\\
&-\frac{1}{2\pi}\int_{\theta_0}^{2\pi-{\theta_0}} d\theta A_\theta(\theta,2\pi) -\frac{1}{2\pi}\int_0^{2\pi} \frac{d\Phi}{2\pi} A_\Phi ({\theta_0},\Phi).
\end{align}
In our case, $A_\theta$ is independent of $\Phi$ so the terms involving it cancel and we have
\begin{align}
C_1 =& \int_0^{2\pi} \frac{d\Phi}{2\pi} [A_\Phi (2\pi-{\theta_0},\Phi) -A_\Phi ({\theta_0},\Phi)].
\end{align}
Similarly computing the integral for the other cylinder we obtain
\begin{align}
C_2 =& \int_0^{2\pi} \frac{d\Phi}{2\pi} [A'_\Phi (2\pi+{\theta_0},\Phi) -A'_\Phi (2\pi-{\theta_0},\Phi)].
\end{align}
The Chern number is then $C=C_1+C_2$. Noting again that $A'_\Phi(\theta)$ is continuous in $S_2$ taking the limit of $\theta_0\rightarrow 0^+$ makes $C_2$ vanish and the Chern number is given by
\begin{align}
C =& \int_0^{2\pi} \frac{d\Phi}{2\pi} [A_\Phi (2\pi^-,\Phi) -A_\Phi (0^+,\Phi)],
\end{align}
or
\begin{align}
C(m)= \tpo(m,\kappa=0^+)-\tpo(m,\kappa=0^-),
\end{align}
where $C(m)$ is the Chern number of any loop around the degenerate point crossing $\kappa=0$ at $m$ and another point in the trivial region ($m>1$).
Using the symmetry of the polarization described above (see the discussion below Eq.~\eqref{eq:sym}) gives
\begin{align}
C(m)= 2\tpo(m,\kappa=0^+)-L,
\label{eq:Chern_number}
\end{align}
which results in $+1$ when $m<1$ and $0$ when $m>1$, as it is known for the SSH chain. The only contribution to the Chern number comes from the discontinuity of $\tpo$ at the branch cut.

\subsection{$\nu = 2$ and disordered case. }

The analysis we did above of $\tpo$ for the SSH chain seems rather trivial as $\tpo$ does not seem to have more information than $\tilde{\mathcal{P}}$ modulo 1. This is no longer true when $t' \neq 0$ allowing for a $\nu = 2$ phase. In Fig.~\ref{ssh_chern}(b) we see $\tpo$ for the dotted cut in the phase diagram with an added symmetry breaking $\kappa$ term. We can see that $\tpo$ is continuous for any path avoiding the branch cut even though the range of values for $\tpo$ is larger than 1. We can see that a loop around the two degenerate points will have a Chern number of $C=2$. 

Jumps in $\tpo$ originate from zero modes changing their occupation. 
Thus, a jump of $n$ in $\tpo$ implies the existence of \emph{at least} $n$ zero-energy modes at that point. 
For this model this equivalence is exact and the jump of $\tpo$ at $\kappa = 0$ gives the number of topological zero-modes, i.e. the winding number. 
Note, however, that there are many ways of breaking the chiral symmetry and for some of them, $\tpo$ may not be related to the winding number any longer. 
For instance, replacing $\kappa$ in Eq.~\eqref{bdi_model} by $\kappa_i=\kappa (-1)^i$, the zero-energy modes split in such a way that $\tpo$ is continuous at $\kappa=0$. It cannot differentiate between the phases $\nu=0$ and $\nu=2$, and a loop like the one discussed above results in a Chern number $C=0$.

To illustrate the advantage of $\tpo$ we also show $\tilde{\mathcal{P}}$ mod $1$ in the $(m,\kappa)$ parameter space in figure~\ref{ssh_chern}(c). As mentioned above the only contribution to the Chern number comes from discontinuities in the bulk polarization.
However, when it is computed using $\tilde{\mathcal{P}}$ we no longer know in advance where this discontinuities are located, due to it being defined modulo 1. One needs to look at the full path to know how the bulk polarization evolves along the loop. 

\begin{figure}[t]
\includegraphics[width=86mm]{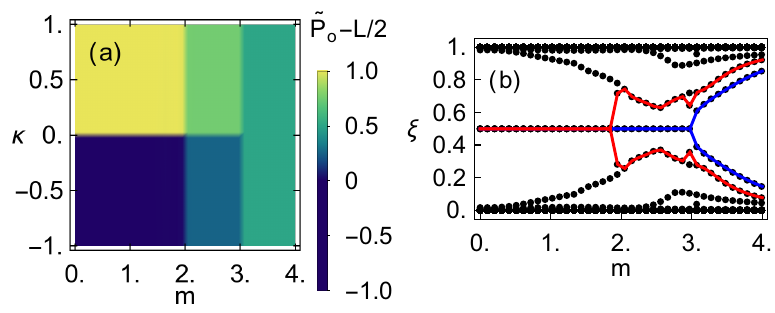}
\caption{(a) Bulk polarization $\tpo$ computed for $t=1,t'=2,\kappa'=0$, with an added disorder to $t$ and $m$ with strength $W=3$, and $\kappa$ is applied only at the edges. We compute it for $L=400$, as the disorder increases the finite size  effect. The disorder opens a region with $\nu=1$ as it has been observed in \cite{Song2014}. (b) EOS for the same parameters as (a), with $\kappa=0$, where we have highlighted the 4 eigenvalues closest to $\xi = 1/2$ to show the phase transitions. As opposed to the energy spectrum, the gap in the EOS does not fill with states such that the virtual topological edge states are still easily identifiable. We show the two eigenvalues closest to $1/2$ in blue, and the next two in red.f }
	\label{disorder_chern}
\end{figure}

Since $\tpo$ is defined in position space we can also use this method for a disordered system. Assume now that we add disorder in the $t$ and $m$ parameters
\begin{align}
&t_i = t + \frac{W}{2} \omega_i \nonumber\\
&m_i = m + W \omega'_i
\end{align}
where $\omega_i$ and $\omega_i'$ are selected from a uniform distribution with range $[-1/2,1/2]$. For very strong disorder where the gap fills with bulk states, $\tpo$ will present many branch cuts as we approach the limit $\lim_{\kappa \rightarrow 0}$, which makes it impractical. In order to avoid this, we break chiral symmetry locally in the edge of the open chain so it only affects the edge modes. If any of the other states cross $E=0$ they do it in  chiral-symmetric pairs such that $\tpo$ is not affected by it. We show $\tpo$ in Fig.\ref{disorder_chern}(a) using the local symmetry breaking term for a strong disorder, $W=3t$, where the gap is indeed filled with states. We see that it works just as in the case without disorder, where now due to the disorder a region of $\nu=1$ opens up \cite{Song2014}. The main difference is that since we break chiral symmetry only for the edge modes, $\tpo$ is quantized to multiples of $1/2$. Note that, since the states that fill the gap are localized in the bulk, they will not appear in the EOS and one can still easily identify the appearance of virtual topological edge states (see figure~\ref{disorder_chern}(b)).
In contrast, the topological edge states in the energy spectrum are completely masked by the (localized) bulk modes that fill the gap.

\subsection{Chern Number in 2D}

We now discuss  the computation of Chern numbers in 2D systems with $\tpo$. Consider the Chern insulator with inversion symmetry given by the Hamiltonian
\begin{align}
H =& \sum_{i\alpha,j\beta}\sum_{k_y} c_{i\alpha}^\dagger(k_y) H_{i\alpha,j\beta}(k_y) c_{j\beta}(k_y), \\ \nonumber
H_{i\alpha,j\beta}(k_y)=& \frac{1}{2}(i\sigma_x-\sigma_z)\delta_{i,j+1}+\frac{1}{2}(-i\sigma_x-\sigma_z)\delta_{i,j-1} \nonumber\\
+&(\sin(k_y)\sigma_y+[2-m-\cos(k_y)]\sigma_z) \delta_{ij}.
\label{eq:chern_model}
\end{align}

\begin{figure}[t]
\centering
\includegraphics[width=86mm]{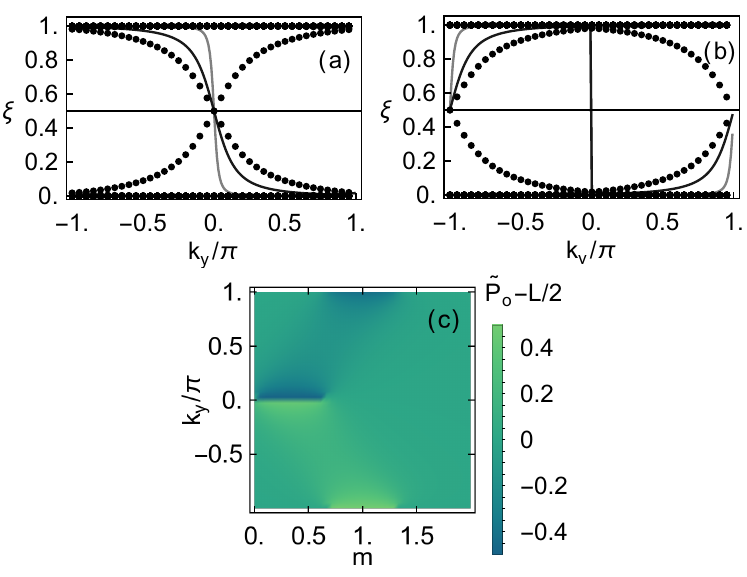}
\caption{(a) and (b) EOS of the Chern insulator in the $C=1$ ($m=1$) and $C=-1$ ($m=3$) phases, respectively. In gray and light gray lines we plot two intermediate points of the adiabatic process of opening the chain between sites $1$ and $L$, showing how the midgap state of the right virtual cut evolves towards the bulk bands, introducing a discontinuity in momentum. (c) $\tpo-L/2$ . Computed for the Chern insulator given by the model in Eq.~\eqref{eq:chern_model} for $L=40$ sites. }
	\label{chern_insulator}
\end{figure}

We can treat the Hamiltonian with elements $H_{i\alpha,j\beta}(k_y)$ as a 1D Hamiltonian with an extra parameter $k_y$ and compute $\tpo$, or equivalently $A_\Phi(\Phi,k_y)$, as in last section. But first lets analyze the EOS of the system with periodic boundary conditions, shown in Fig.~\ref{chern_insulator}(a) and (b) for $m=1$ and $3$, respectively. For $0<m<2$ the EOS presents two edge modes that connect the occupied and empty bands that cross $\xi=1/2$ at $k_y=0$. In this region the occupied bands have a Chern number $C=+1$. For $2<m<4$ there are also edge modes but they cross $\xi=1/2$ at $k_y=\pi$ and the system has $C=-1$. For $m<0$ or $m>4$ the system is a trivial insulator with $C=0$. When we open the chain, the eigenvalue of the right virtual topological edge mode will be pushed to $\xi=0$ or $1$. This is seen in figure~\ref{chern_insulator}(a) and (b) for intermediate points in the process of opening the chain. This eigenvalue presents a jump whenever the correspondent edge-mode crosses zero energy, going between $\xi=1$ and $\xi=0$, i.e. the number of eigenvalues at $\xi = 1$ is not constant across $k_y$. In this model, due to inversion symmetry, this is only possible at $k_y=0,\pi$. 

We can now take a look at $\tpo$ in the parameter space $(m,k_y)$, shown in figure~\ref{chern_insulator}(c). We see, again, that $\tpo$ remains continuous except for the branch cuts that appear at the points where the physical edge-modes cross zero energy. The Chern number can be obtained as
\begin{equation}
C(m) = \int_{-\pi}^{\pi}\frac{dk_y}{2\pi} \partial_{k_y} \tpo(m,k_y).
\label{eq:Zaletel_chern}
\end{equation}
However, since we know the position of the branch cuts we can integrate avoiding them and obtain
\begin{align}
C(m) =& [\tpo(m,k_y=0^-)-\tpo(m,k_y=-\pi^+)] \nonumber \\ 
&+[\tpo(m,k_y=\pi^-)-\tpo(m,k_y=0^+)]. 
\end{align}
The expression in the right hand side is actually a known topological invariant, the trace index \cite{Alexandrinata2011}, defined in terms of the trace of $C_{A,o}$. In reference \cite{Alexandrinata2011} it is shown to be equal to the Chern number by relating it to the Hall current. Here, the connection between the trace index and the Chern number is seen by framing $\Tr[C_{A,o}]$ as a geometric quantity


\section{Conclusion}

It was previously known that the bulk polarization (Zak phase, or geometric phase for $U(1)$ flux insertion) was encoded in the ES. Here we have shown how it can be obtained from the single-particle ES of a non-interacting gapped system, even when it is not quantized, showing that there is substantially more information in the single-particle ES than previously realized. 
Our formulation is both simpler than the one involving the ES, which is much more difficult to compute, and it provides additional insight into the single-particle ES and EOS. In particular in the topological case, the relation between the quantized bulk polarization (or the Chern number in 2D) and the number of virtual topological edge states is demonstrated in a particularly transparent way. It also provides a new simple method for computing the bulk polarization for systems without translational invariance. 

We also define a novel bulk polarization that is continuous in $\mathbb{R}$ for gapped paths. This provides additional information about the edge spectrum, similar to how the Zak phase is equal to the winding number when computed in a particular gauge. Our new bulk polarization simplifies the calculation of changes in the bulk polarization, and provides a new route to the computation of Chern numbers.

\emph{Outlook:} In the application of our results to topological systems in 1D we have focused on systems with translational invariance even though our results do not require it. For general spatially inhomogeneous systems the same bulk can support different boundaries so the relation between bulk and boundary is not as straightforward as in the translational invariant case. Our results might give more insight into this relation. 

As it stands, the relation derived in this paper only works for non-interacting systems. With interactions, the correlation matrix does not have the full information of the ground state. However, there might be certain systems, perhaps for weakly interacting systems, where the correlation matrix encodes the topological information. This would be worth exploring since, even for interacting systems, the correlation matrix is much easier to compute than the density matrix.

In the field of higher-order topological insulators some systems can be characterized by a physical generalization of the bulk polarization to two dimensions, the quadrupole moment \cite{Benalcazar2017,Kang2019}. This quadrupole moment is defined via the position operator and it is an open question whether there is a formulation in terms of twisted boundary conditions. In this case, one might be able to construct this quadrupole moment using the ES, as we did
 here for the bulk polarization.

\acknowledgments
{\em Acknowledgments.--} 
The research in this grant was supported by the Swedish Research Council under grant no. 2017-05162 and the Knut and Alice Wallenberg foundation under grant no. 2017.0157. 
We thank Krishanu Roychowdhury and Thors Hans Hansson for helpful discussions and a critical reading of the manuscript. 

\bibliography{references}

\appendix


\section{Computing the Bulk polarization from the Schmidt decomposition}\label{app:pollmann}

In this appendix we review  how to compute the bulk polarization from the Schmidt decomposition~\cite{Zaletel2014} and relate it to our own results.

Consider a closed chain of length $L$ and a bipartition into regions $A$, for $i\in [1,L/2]$, and $B$, for $i \in [L/2+1,L]$. Using a Schmidt decomposition the ground state gives
\begin{equation}
\ket{\Psi^0} = \sum_{p,q} s_p s'_q \ket{p,q}_A \ket{q,p}_B,
\end{equation}
The Schmidt indices $p$ and $q$ label the fluctuations at the left ($i=1$) and right ($i=L/2$) cuts, respectively. The convention used for the states $\ket{p,q}_{R}$ is that the first (second) index labels the state near the left-most (right-most) region of $R$, where $R$ can be $A$ or $B$. Assuming the system is large enough compared with the correlation length the fluctuations across the two cuts are independent of each other. The reduced density matrix can be computed as
\begin{equation}
\rho_A = \sum_{p,q} s_p^2 s^{\prime \, 2}_q \ket{p,q}_{A}\bra{p,q}_A .
\end{equation}
As mentioned in section \ref{section:ES} it is fully determined by the correlation matrix $C_A$. In the state $\ket{p,q}_A$, the Schmidt index $p$ ($q$) labels a set of occupation numbers of the eigenstates of $C_A$ related to the cut at $i=1$ ($i=L/2$). we will refer to these subspaces of the Hilbert space as $A_L$ ($A_R$). Note that the remaining subspace $A_{\rm bulk}$ is composed of eigenstates that have eigenvalues exponentially close to $\xi = 0,1$. The eigenvalues of $\rho_A$, $\lambda_{pq}=s_p^2 s_q^{\prime \, 2}$ can be obtained as \cite{Alexandrinata2011}
\begin{align}
\lambda_{pq} =& \prod_{\mu \in A} (1-\xi_\mu)\left(\frac{\xi_\mu}{1-\xi_\mu}\right)^{n_\mu^{pq}}\nonumber \\
=&\left(\prod_{\mu\in A_L} (1-\xi_\mu)\left(\frac{\xi_\mu}{1-\xi_\mu}\right)^{n_\mu^{p}} \right) \nonumber\\
&\cross \left(\prod_{\nu \in A_R} (1-\xi_\nu)\left(\frac{\xi_\nu}{1-\xi_\nu}\right)^{n_\nu^{q}} \right) ,
\label{eq:Alexandrinata}
\end{align}
from which we identify 
\begin{align}
&s_p^2 = \prod_{\mu\in A_L} (1-\xi_\mu)\left(\frac{\xi_\mu}{1-\xi_\mu}\right)^{n_\mu^{p}} \nonumber \\
&s_q^{\prime 2} = \prod_{\nu \in A_R} (1-\xi_\nu)\left(\frac{\xi_\nu}{1-\xi_\nu}\right)^{n_\nu^{q}}.
\label{eq:spsq_A}
\end{align}

We are now in position to introduce the $U(1)$ flux via a twisted boundary condition as
\begin{equation}
\ket{\Psi^\Phi} = \sum_{pq} s_p s_q' e^{-i\Phi N_{A_L}^p} \ket{p,q}_A \ket{q,p}_B,
\label{eq:Ground_state_flux}
\end{equation}
where $N_{A_L}^p = \sum_{\mu \in A_L}n_\mu^p$. Note that there is an ambiguity in how the flux is introduced as one can always include bulk modes into $A_L$. As mentioned above, the label $p$ describes the set of occupation numbers $\{n_\mu^p\}$. Equation \eqref{eq:Ground_state_flux} means that an electron crossing the virtual cut at $i=1$ between region $B$ and $A_L$ will acquire a phase $-\Phi$, the convention on the sign is the same as the one used in reference \cite{Watanabe2018}. 

The polarization can be now computed as
\begin{align}
\tilde{\mathcal{P}} =& \int_0^{2\pi} \frac{d\Phi}{2\pi}i\bra{\Psi^\Phi}\partial_\Phi\ket{\Psi^\Phi} \nonumber\\
=& \int_0^{2\pi} \frac{d\Phi}{2\pi}i\sum_{p,q} s_p^2 s_q^{\prime \, 2} e^{i\Phi N_{A_L}^p}\partial_{\Phi}e^{-i\Phi  N_{A_L}^p}\nonumber\\
=&\int_0^{2\pi} \frac{d\Phi}{2\pi}\sum_{p,q} s_p^2 s_q^{\prime \, 2}  N_{A_L}^p \nonumber\\
=&\sum_{p} s_p^2 \sum_{\mu \in A_L}n^p_\mu,
\end{align}
where we used that $\sum_q s_q^{\prime \,2} = 1$. Inserting now the expression for $s_p^2$ we obtain
\begin{align}
\tilde{\mathcal{P}}=&\sum_{\{n_{i \in A_L}\} = 0,1} \left(\sum_{j\in A_{L}} n_j \right) \left(\prod_{k\in A_L} \lambda_k \right),
\end{align} 
where we defined
\begin{equation}
\lambda_k = (1- \xi_k)\left(\frac{\xi_k}{1-\xi_k} \right)^{n_k}.
\end{equation}

If we expand the sum for one particular occupation number $n_p$ we have
\begin{align}
\tilde{\mathcal{P}} =& \sum_{\substack{\{n_{i \neq p \in A_L}\} = 0,1 \\ n_p=0}} \left(\sum_{j\neq p \in A_{L}} n_j \right) (1-\xi_p)\left(\prod_{k \neq p\in A_L} \lambda_k \right) \nonumber \\
&+ \sum_{\substack{\{n_{i \neq p \in A_L}\} = 0,1 \\ n_p=1}} \left(\sum_{j\neq p \in A_{L}} n_j+1 \right)\xi_p \left(\prod_{k \neq p\in A_L} \lambda_k \right) \nonumber \\
=& \sum_{\substack{\{n_{i \neq p \in A_L}\} = 0,1 }} \left(\sum_{j\neq p \in A_{L}} n_j \right) \left(\prod_{k \neq p\in A_L} \lambda_k \right) \nonumber \\
&+ \xi_p \sum_{\substack{\{n_{i \neq p \in A_L}\} = 0,1 }} \left(\prod_{k \neq p\in A_L} \lambda_k \right) 
\end{align}
If we continue expanding the sum in the first term we will get terms like the second one for all the other $k\neq p$ eigenvalues. If we further expand the sum in the factor accompanying $\xi_p$ we have
\begin{align}
&(1-\xi_{p'})\sum_{\substack{\{n_{i \neq p,p' \in A_L}\} = 0,1 \\ n_{p'}=0}} \left(\prod_{k \neq p,p'\in A_L} \lambda_k\right) \nonumber \\
&+ \xi_{p'}\sum_{\substack{\{n_{i \neq p,p' \in A_L}\} = 0,1 \\ n_{p'}=1}} \left(\prod_{k \neq p,p'\in A_L}\lambda_k\right) =1.
\end{align}
If we continue this procedure for all other occupation numbers we arrive at
\begin{align}
\tilde{\mathcal{P}} =& \sum_{p\in A_L}\xi_p  \quad {\rm mod}\, 1.
\end{align}
In terms of the spectrum of $C_A$ the bulk polarization simplifies greatly. As mentioned above, the subspace $A_L$ is not well-defined, as one can always include bulk modes, however since the bulk polarization is defined modulo 1, this issue is irrelevant. In practice we extend $A_L$ to include all eigenstates whose average position lies in the left half of $A$ (which we denote by $L$) and we finally get
\begin{align}
\tilde{\mathcal{P}} =& \sum_{i\in L}\xi_i \quad {\rm mod}\, 1.
\label{eq:polarization_left}
\end{align}

\section{Alternative expression for the correlation matrix}\label{app:CM}
In this appendix, we show how to express the correlation matrix in terms of the Hamiltonian, used in Eq.~\eqref{eq:corr_mat2}. 
We consider the generic, quadratic Hamiltonian in one dimension of Eq.~\eqref{eq:quadr_Ham}, which is diagonalized by a unitary matrix $U$ with 
	\begin{align}\label{eq:D}
	H&=UDU^\dagger& \mbox{with } D=\mbox{diag}(E_{p\mu}).
	\end{align} 
In terms of the fermionic operators that diagonalize the Hamiltonian, 
\begin{align}
& \gamma_{i\alpha} = \sum_{j\beta}\psi_{j\beta}^{i\alpha \, \ast} c_{j\beta},
\end{align}
we find that the correlation matrix can be written as 
\begin{align}\label{eq:corr_mat}
C_{ij}^{\alpha \beta} =& \sum_{pq,\mu\nu} \psi^{q\nu }_{j\beta} \psi^{p\mu \, \ast}_{i\alpha} \expval{\gamma^\dagger_{p\mu} \gamma_{q\nu} }\nonumber \\
=&  \sum_{p\mu} \psi^{p\mu }_{j\beta} \psi^{p\mu \, \ast}_{i\alpha} \expval{\gamma^\dagger_{p\mu} \gamma_{p\mu} }\nonumber \\
=&  \sum_{p\mu} \psi^{p\mu }_{j\beta} \psi^{p\mu \, \ast}_{i\alpha} [1-{\rm sign}(E_{p\mu})]/2.
\end{align}
The first term of the last line is simply a kronecker delta between both sets of indices. The second term can be rewritten, using Eq.~\eqref{eq:D}, as
\begin{align}
UD(\abs{D})^{-1}U^\dagger =& UDU^\dagger U(D^2)^{-1/2}U^\dagger
\end{align}
We can rewrite this expression further by noting that 
\begin{align}
[U (D^2)^{-1/2} U^\dagger]^2 =& U (D^2)^{-1/2} U^\dagger U  (D^2)^{-1/2} U^\dagger \nonumber\\
=&  U (D^2)^{-1/2}  (D^2)^{-1/2} U^\dagger \nonumber\\
=&  U (D^2)^{-1} U^\dagger \nonumber\\
=&  (U D^2 U^\dagger)^{-1}.
\end{align}
Therefore,
\begin{align}
U (D^2)^{-1/2} U^\dagger =&(U D^2 U^\dagger)^{-1/2} 
\end{align}
and we conclude that
\begin{align}\label{eq:UDDU}
UD(\abs{D})^{-1}U^\dagger =& UDU^\dagger(U D^2 U^\dagger)^{-1/2} \nonumber\\ 
=& H(H^2)^{-1/2}.
\end{align}
Combining Eq.s~\eqref{eq:corr_mat},  and \eqref{eq:UDDU}, we arrive at the final expression of the correlation matrix in Eq.~\eqref{eq:corr_mat2}. 

\section{Winding number in the SSH chain}
\label{appendix:winding_ssh}

\begin{figure}[t]
	\centering
	\includegraphics[width=50mm]{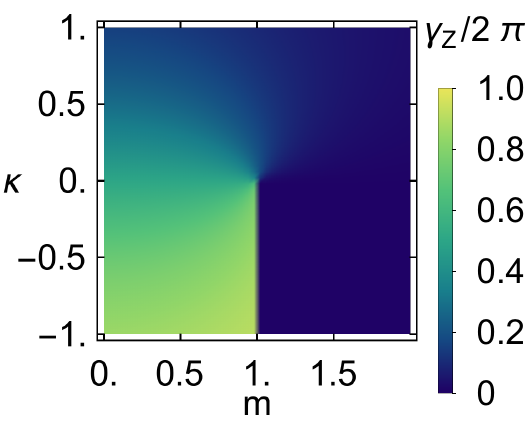}
	\caption{$\gamma/2\pi$ obtained in the smooth gauge that provides the winding number in the chiral symmetric limit.}
	\label{fig:ssh_zak}
\end{figure}

Consider the Bloch Hamiltonian of the Rice-Mele model
\begin{equation}
H=\mqty(\kappa & f(k) \\ f^\dagger(k) & -\kappa),
\end{equation}
which can also be written as $H = \boldsymbol{h}\cdot \boldsymbol{\sigma}$
where 
\begin{align*}
\boldsymbol{h}=& ({\rm Re}[f(k)],-{\rm Im}[f(k)],\kappa) \\
=& \sqrt{\abs{f(k)}^2 +\kappa^2}(\sin(\theta)\cos(\phi),\sin(\theta)\sin(\phi),\cos(\theta))
\end{align*}
In the gauge where the second component of the eigenstates remains real they are given by
\begin{align}
&\ket{+}  = \frac{1}{N_+}\mqty(\cot(\theta/2)e^{-i\phi}\\ 1 )\\
&\ket{-}  =\frac{1}{N_-} \mqty(-\tan(\theta/2)e^{-i\phi}\\ 1),
\end{align}
where
\begin{align*}
&N_+ = \sqrt{1+\cot^2(\theta/2)} \\
&N_- = \sqrt{1+\tan^2(\theta/2)} .
\end{align*}
The Berry connections of the occupied state, $A_\lambda = \bra{-}\partial_\lambda \ket{-}$, can be obtained as
\begin{align}
A_{\lambda}=& i\frac{\cos(\theta)-1}{2}\partial_q \phi \\
=&i\frac{\cos(\theta)-1}{2} \cos(\phi)^2 \partial_q  \tan(\phi) \nonumber \\
=&\frac{i}{2}\left(\frac{\kappa}{\sqrt{\abs{f(k)}^2+\kappa^2}}-1\right) \left( \frac{\Re [f(k)]}{\abs{f(k)}}\right)^2 \partial_\lambda  \left( -\frac{\Im [f(k)]}{\Re [f(k)]} \right) \nonumber \\
=&\frac{-1}{2}\left(\frac{\kappa}{\sqrt{\abs{f(k)}^2+\kappa^2}}-1\right) \frac{f(k)^\dagger \partial_\lambda f(k)-f(k)\partial_\lambda f(k)^\dagger}{2\abs{f(k)}^2} \nonumber
\end{align}
The Berry connection with respect to momentum gives
\begin{equation}
A_k = \frac{it [t+m \cos(k)][\kappa - \sqrt{\kappa^2+m^2+t^2+2mt \cos(k)}]}{2[m^2+t^2+2mt\cos(k)]\sqrt{\kappa^2+m^2+t^2+2mt \cos(k)}},
\end{equation}
and the resulting Zak phase can be seen in Fig.~\ref{fig:ssh_zak}, which presents a branch cut for $m=0,\kappa<0$ for this particular gauge.

Note that in the chiral symmetric case, for $\kappa = 0$, the Zak phase is
\begin{align}
\gamma =& \int_0^{2\pi} dk\,\frac{1}{2}\frac{f(k)^\dagger \partial_k f(k)-f(k)\partial_k f(k)^\dagger}{2\abs{f(k)}^2} \nonumber \\
=& \int_0^{2\pi} dk\,\frac{1}{2}q(k)^\dagger \partial_k q(k),
\end{align}
where $q(k) = f(k)/\abs{f(k)}$. This is nothing else than the winding number \cite{ryu2010topological}. 

\end{document}